\documentclass[conference]{IEEEtran}
\pdfoutput=1

\usepackage{etoolbox}
\newbool{isgitdraft}
\boolfalse{isgitdraft}

\usepackage{xcolor}
\usepackage[normalem]{ulem}
\newbool{isReviewUpdate}
\boolfalse{isReviewUpdate}

\usepackage{cite}

\usepackage{multirow}

\usepackage[pdftex]{graphicx}
\graphicspath{{imgs/}}
\DeclareGraphicsExtensions{.pdf,.jpeg,.png}

\usepackage{amsmath}
\usepackage{amsfonts}
\usepackage{amsthm}
\usepackage{amssymb}
\usepackage{mathtools}

\usepackage{algpseudocode}
\usepackage{algorithm}

\algnewcommand{\LeftComment}[1]{\Statex \(\triangleright\) #1}

\usepackage{array}

\usepackage[caption=false,font=footnotesize]{subfig}

\usepackage{url}

\ifbool{isgitdraft}{
  \usepackage[mark]{gitinfo2}
}

\usepackage{styles/rfs_acros}

\usepackage{styles/rfs_macros}

\begin{document}

\ifbool{isgitdraft}{
  \begin{titlepage}
    \centering
    \vspace*{5cm}
    \huge\textit{\textbf{DRAFT}}: Efficient Implementation of Multi-sensor Adaptive Birth Samplers for Labeled Random Finite Set Tracking \\
    \vspace{2\baselineskip}
    \large Compiled on \today\\
    \vspace{2\baselineskip}
    \textbf{Git Hash}: \texttt{\gitHash}\\
    \vspace{2\baselineskip}
    \textbf{Git Branch}: \texttt{\gitBranch}\\
    \vfill
  \end{titlepage}
  \newpage
}

\title{Efficient Implementation of Multi-sensor Adaptive Birth Samplers for Labeled Random Finite Set Tracking}

\author{
  \IEEEauthorblockN{
    Jennifer Bondarchuk,
    Anthony~Trezza,
    Donald~J.~Bucci~Jr.,
  }
  \IEEEauthorblockA{
    \textit{Advanced Technology Laboratories} \\
    \textit{Lockheed Martin} \\
    Cherry Hill, NJ, USA \\
    \{jennifer.a.bondarchuk, anthony.t.trezza, donald.j.bucci.jr\}@lmco.com
  }
}

\maketitle
\begin{abstract}
    Adaptive track initiation remains a crucial component of many modern multi-target tracking systems.
    For labeled random finite sets multi-object filters, prior work has been established to construct a labeled multi-object birth density using measurements from multiple sensors.
    A naive construction of this adaptive birth set density results in an exponential number of newborn components in the number of sensors.
    A truncation procedure was provided that leverages a Gibbs sampler to truncate the birth density, reducing the complexity to quadratic in the number of sensors.
    However, only a limited discussion has been provided on additional algorithmic techniques that can be employed to substantially reduce the complexity in practical tracking applications.
    In this paper, we propose five efficiency enhancements for the labeled random finite sets multi-sensor adaptive birth procedure.
    Simulation results are provided to demonstrate their computational benefits and show that they result in a negligible change to the multi-target tracking performance.
\end{abstract}


\IEEEpeerreviewmaketitle
\section{Introduction}\label{sec::intro}
The objective of a \ac{MTT} algorithm is to jointly estimate the number of objects in the scene and their trajectories from noisy measurements observed at one or more sensors over time.
\ac{MTT} solutions are often required in a variety of applications domains such as defense, automotive, and maritime traffic monitoring systems  \cite{Blackman1999}.
Many approaches have been proposed including \ac{GNN} techniques~\cite{Blackman1999},~\ac{JPDA}~\cite{Barshalom2009},~\ac{MHT}~\cite{Blackman2004},~\ac{BP}~\cite{Meyer2018}, and~\ac{RFS}~\cite{Mahler2007, Mahler2014}.
We direct the reader to~\cite{Vo2015} for a detailed survey of the field, recent advances, and example applications.
In this work, we specifically focus on the labeled \ac{RFS} problem formulation of the $\delta$-\ac{GLMB} and \ac{LMB} filters~\cite{Vo2013, Vo2014, Reuter2014, Reuter2017}.

It has been shown that the $\delta$-\ac{GLMB} labeled \ac{RFS} density is a conjugate prior with itself under the multi-object Bayes filtering equations~\cite{Vo2013,Vo2014,Vo2019}, resulting in a tractable solution to the \ac{MTT} problem.
In the $\delta$-\ac{GLMB} filter, the predicted labeled multi-object state is modeled as the superposition of surviving and newborn targets.
Under the assumption that targets move, appear, and die independently of each other when conditioned on the prior state, the transition density for a labeled multi-object state becomes the product of a transition density over the surviving targets and a prior density over the newborn targets in the predicted state.
An inaccurate model for the newborn target density can result in missing targets, increasing the number of ghost tracks, increasing the runtime complexity, and decreasing the tracker accuracy.

In an adaptive birth strategy, the newborn target density is constructed adaptively from the measurements at each time step~\cite{Ristic2012}.
This approach is effective in many applications since minimal prior information is known about where and how objects can appear in the surveillance volume.
Care must be taken when designing these strategies to ensure that tracker performance is maintained without adversely affecting computational complexity.
Single-sensor adaptive birth techniques for labeled \ac{RFS} were introduced in~\cite{Reuter2014} for the~\ac{LMB} filter and~\cite{Lin2016} for the~$\delta$-\ac{GLMB} filter.
These approaches follow a similar concept as presented in~\cite{Ristic2012} by constructing a component of an \ac{LMB} RFS from every single-sensor measurement observed at the previous time step.
The probabilities of birth for each of these \ac{LMB} components were set proportionally to a maximum birth rate parameter.
The proportionality factor was based on how much a measurement did not associate with the persisting targets from the last time step.
In contrast, the multi-sensor adaptive birth regime carries the additional complication that there are exponentially many possible measurement tuples that could be used to generate newborn targets.
In our previous work~\cite{Trezza2022}, we proposed a stochastic Gibbs sampling approach for truncating multi-sensor measurement tuples that would likely be pruned in the subsequent $\delta$-\ac{GLMB} update step.
The proposed Gibbs sampler exhibited quadratic complexity in the number of sensors, and both an \ac{SMC} approximation and a closed form Gaussian solution were provided.
In~\cite{Trezza2023}, we proposed an alternative Gibbs sampling architecture and two early termination criteria that can reduce the number of Markov chain observations that the Gibbs sampler must consider, but focused the analysis primarily on the $\delta$-\ac{GLMB} ranked assignment problem.

As discussed in \cite{Trezza2023}, despite the approach in \cite{Trezza2022} having quadratic scaling complexity, each Gibbs iteration can contribute to a non-negligible amount of the filter’s runtime.
For example, in our simulations presented in Section~\ref{sec::sim}, if no efficient methods are used, the adaptive birth procedure in \cite{Trezza2022} accounts for 98\% of the simulation's runtime.
Hence, any runtime improvements made to adaptive birth contribute similarly to the overall runtime reduction.

In this paper, we expand the work of~\cite{Trezza2022} by proposing five practical techniques to improve the runtime efficiency, including pre-pruning, gating, memoization, missed detection sample skipping, post-pruning and capping.
This paper is organized as follows.
Section~\ref{sec::background}, background on adaptive birth.
Section~\ref{sec::proposed} details the proposed efficiency methods.
Section~\ref{sec::sim} provides simulation results for each proposed method, followed by conclusions in Section~\ref{sec::conclusions}.

\section{Background}\label{sec::background}
We adopt the notation of~\cite{Vo2013, Vo2019, Trezza2022} as follows.
Lowercase letters denote vectors (e.g., $x, \textbf{x}$), whereas uppercase letters denote finite sets (e.g., $X, \textbf{X}$).
Bold letters denote labeled states and their distributions (e.g., $\textbf{x}, \textbf{X}, \boldsymbol{\pi}$) and blackboard bold letters denote spaces (e.g., $\mathbb{X}, \mathbb{Z}, \mathbb{L}, \mathbb{R}$).
Variable sequences $X_i, X_{i+1}, \dots, X_j$ are abbreviated $X_{i:j}$.
The inner product $\int f(x) g(x) dx$ is written as $\langle f, g \rangle$.
Finally, we drop the subscript notation for current time step $k$ and use subscript '$+$' to indicate the next time step ($k+1$).

We assume the standard point-object models described in~\cite{Vo2013} that result in a closed-form $\delta$-\ac{GLMB} filtering recursion.
In this formalism, the newborn birth density is modeled as a labeled \ac{RFS} density.

\subsection{Multi-sensor Multi-target Adaptive Birth Model}
Our objective is to adaptively construct an \ac{LMB}~\ac{RFS} density $\textbf{f}_{B,+}$ given the multi-object measurement sets $Z^{(s)} \in \mathbb{Z}^{(s)}$ from sensors $s \in \{1, \dots, V\}$.
Let $m^{(s)} = |Z^{(s)}|$ and $\mathbb{J}^{(s)} \triangleq \{1, \dots, m^{(s)}\}$ be an index set into $Z^{(s)}$.
Let $\mathbb{J}^{(s)}_0 \triangleq \mathbb{J}^{(s)} \cup \{0\}$, where the entry $0$ represents a miss-detection.
The tuple $J \triangleq (j^{(1)}, \dots, j^{(V)}) \in \mathbb{J}^{(1)}_0 \times \dots \times \mathbb{J}^{(V)}_0$ is a multi-sensor measurement index with $Z_J \triangleq (z^{(1)}_{j^{(1)}}, \dots, z^{(V)}_{j^{(V)}})$ as the corresponding tuple of multi-sensor measurements.

As suggested in~\cite{Trezza2022}, the birth \ac{LMB}~\ac{RFS} density used at time ($k+1$) is modeled as
\begin{equation}
    \textbf{f}_{B,+} = \left\{ \left(r_{B,+}(l_+), p_{B,+}(\cdot, l_+ | Z_J) \right) \right\}_{l_+ \in \mathbb{B}_+},
\end{equation}
where $l_+ \in \mathbb{B}_+ = \{(k+1, J) : \forall J \in \mathbb{J}_0\}$ is the newborn object's label having birth probability $r_{B,+}(l_+)$.
The quantity $p_{B,+}(\cdot, l_+ | Z_J)$ is the state distribution of label $l_+$ conditioned on the measurement tuple $Z_J$.
The state distribution is given by the Chapman-Kolmogrov equation \cite{Mahler2007}
\begin{equation}\label{eq::pred_birth_post}
    p_{B,+}(x_+, l_+ | Z_J) = \int f_+(x_+|x, l) p_B(x, l_+ | Z_J) dx,
\end{equation}
where,
\begin{align}
    p_B(x, l_+ | Z_J)       & = \frac{p_B(x, l_+) \psi^{J}_Z(x, l_+)}{\bar{\psi}^{J}_{Z}(l_+)} \label{eq::spatial_distr} \\
    \bar{\psi}^{J}_{Z}(l_+) & = \langle p_B(\cdot, l_+), \psi^{J}_{Z}(\cdot, l_+)\rangle. \label{eq::psi_bar}
\end{align}
Here $f_+(x_+|x, l)$ is the Markov transition density modeling the dynamics of the object state, $\psi^{J}_Z(x, l_+)$ is the multi-sensor measurement pseudolikelihood function, and $p_B(x, l_+)$ is the prior distribution on the newborn object's state.
Assuming conditional independence of the sensor measurements, the multi-sensor measurement pseudolikelihood function is~\cite{Vo2019},
\begin{equation}\label{eq::joint_likelihood}
    \psi^J_Z(\textbf{x}) \triangleq \prod\limits^V_{s=1} \psi^{s, j^{(s)}}_{Z^{(s)}}(\textbf{x}),
\end{equation}
where $\psi^{s, j^{(s)}}_{Z^{(s)}}(\textbf{x})$ is the per-object, per-sensor measurement pseudolikelihood~\cite{Vo2013, Vo2015},
\begin{equation}\label{eq::meas_likelihood}
    \psi^{s, j^{(s)}}_{Z^{(s)}}(\textbf{x}) =
    \begin{cases}
        \frac{p_D^{(s)}(\textbf{x})g^{(s)}(z^{(s)}_{j^{(s)}} | \textbf{x})}{\kappa^{(s)}(z^{(s)}_{j^{(s)}})} & j^{(s)} \in \mathbb{J}^{(s)} \\
        1 - p_D^{(s)}(\textbf{x})                                                                            & j^{(s)} = 0
    \end{cases}
\end{equation}
and $p_D^{(s)}(\textbf{x})$ models the probability that sensor $s$ detected the target state $\textbf{x}$, $g^{(s)}(z^{(s)}_{j^{(s)}}| \textbf{x})$ is the single-sensor likelihood function and $\kappa^{(s)}(z^{(s)}_{j^{(s)}})$ is the Poisson \ac{RFS} intensity function of the clutter process.

The birth probability of label $l_+$ is modeled in \cite{Trezza2022} as
\begin{equation}\label{eq::birth_prob}
    r_{B,+}(l_+) = \min\left(r_{B, \text{max}}, \hat{r}_{B,+}(l_+)\lambda_{B,+} \right),
\end{equation}
where $r_{B, \text{max}} \in [0, 1]$ is the maximum existence probability of a newborn target and $\lambda_{B,+}$ is the expected number of target births at time step $k+1$.
The effective birth probability $\hat{r}_{B,+}(l_+)$ is given as
\begin{equation}\label{eq::birth_prob_hat}
    \hat{r}_{B,+}(l_+) = \frac{r_U(J) \bar{\psi}^{J}_{Z}(l_+)}{\sum\limits_{J' \in \mathbb{J}_0} r_U(J')\bar{\psi}^{J'}_{Z}(l_+)},
\end{equation}
where $r_U(J)$ is the probability that the multi-sensor measurement tuple $J$ does not associate with any existing targets.
This non-association probability is approximated as
\begin{equation}\label{eq::unassoc_prob}
    r_{U}(J) \propto \prod_{j \in J} \left(1 - r_{A}(j)\right),
\end{equation}
where $r_{A}(j)$ is the probability that a single-sensor's measurement, $z^{(s)}_{j^{(s)}}$, is associated with existing targets in the $\delta$-\ac{GLMB} posterior density.
For more information on these quantities, we direct the reader to~\cite{Trezza2022}.

\subsection{Truncating the Multi-sensor Adaptive Birth LMB}
The number of newborn labels $\mathbb{B}_+$ is driven by the cardinality of $\mathbb{J}_0$ which has $O(m^V)$ elements, where $m$ is the worst case number of measurements.
It is therefore not practical to enumerate every possible entry at each time step.
In the spirit of \cite{Vo2017}, a technique was proposed in \cite{Trezza2022} to sample labels $\mathbb{B}'_+ \subset \mathbb{B}_+$ that have high $\hat{r}_{B,+}(l_+) $ and hence are likely to persist through future measurement updates.
The sampling distribution is modeled as
\begin{equation}\label{eq::orig_sampling_distr}
    p(l_+) \propto \hat{r}_{B,+}(l_+) \propto r_U(J)\bar{\psi}^{J}_{Z}(l_+).
\end{equation}
Directly sampling from the categorical distribution $p(l_+)$ requires evaluation of an exponential number of possible birth labels.
By construction of the birth set, sampling from $p(l_+)$ is equivalent to sampling multi-sensor measurement tuples from the joint distribution, $p(J) = p(j^{(1)}, \dots, j^{(V)})$.
A Gibbs sampler was proposed in \cite{Trezza2022} to achieve efficient sampling from Equation~(\ref{eq::orig_sampling_distr}) using the Markov transition probabilities
\begin{equation}\label{eq::cdn_likelihood}
    p(j^{(s)} | J^{-s}) \propto \left(1 - r_A(j^{(s)})\right) \bar{\psi}^{J}_{Z}(l_+),
\end{equation}
where $J^{-s} = (j^{(1)}, \dots, j^{(s-1)}, j^{(s+1)}, \dots, j^{(V)})$.

The stochastic Gibbs sampler that generates the measurement tuples (i.e., birth components) with significant birth probabilities is given in \cite[Algorithm 1]{Trezza2022}.
Similar to \cite{Vo2017}, it exhibits an exponential convergence rate and does not require a burn-in period since unique solutions can be directly used in construction of the birth set.
An \ac{SMC} method was provided in \cite[Section VI]{Trezza2022} for evaluating the transition probabilities and generating the state distributions.
Closed-form solutions for these quantities in the Gaussian case can be found in \cite[Section VII]{Trezza2022}.


\section{Efficiency Improvements}\label{sec::proposed}
In this section we discuss our proposed methods to reduce the runtime of the adaptive birth procedure.
We describe the motivation for each method as well as our implementation.

\subsection{Associated Measurement Pre-pruning}\label{sec::proposed::preprune}
To pre-prune measurements, we exploit the fact that each measurement's association probability must be computed at the previous time step and provide an efficient check to remove measurements from the sample space that are likely to have been associated with existing targets.

The association probability, $r_A(j^{(s)})$, is computed for every measurement and provided as an input to the multi-sensor adaptive birth procedure.
As seen in Equation~(\ref{eq::unassoc_prob}), the joint unassociation probability assumes conditional independence between sensors.
If any measurement $z^{(s)}_{j^{(s)}}$ in the measurement tuple $j^{(s)} \in J$ has a high association probability, then $r_U(J)$ will be low.
By Equation~(\ref{eq::birth_prob_hat}), a low value for $r_U(J)$ results in a low value for the birth probability in Equation~(\ref{eq::birth_prob}).
This is intuitive as a low unassociation probability indicates that at least one measurement in the measurement tuple is likely to have associated with an existing target and therefore should not be used to birth a newborn target.

This is accounted for by construction of the Markov transition density in Equation~(\ref{eq::cdn_likelihood}) such that if $r_A(j^{(s)})$ is high, then it is unlikely that $j^{(s)}$ will be transitioned to and therefore will be unlikely to be sampled.
However, in the current implementation, the value for $\bar{\psi}^J_Z$ in Equation~(\ref{eq::psi_bar}) still needs to be computed or approximated.

In practice, it has been observed that for highly accurate sensors, $r_A(j^{(s)})$ tends to either be very close to $0$ or $1$.
Additionally, in practical multi-sensor, multi-target tracking applications with many targets in the scene, it is possible that many measurements have a high value for $r_A(j^{(s)})$ and thus results in many unnecessary and computationally costly computations of $\bar{\psi}^J_Z$.

Since the association probability is provided as an input, we propose a simple pre-pruning approach to remove measurements from the sample space whose association probability is above a threshold, $r_A(j^{(s)}) > \tau_A$ with $\tau_A \in [0, 1]$.

\subsection{Measurement Gating}\label{sec::proposed::gate}
In Equation~(\ref{eq::joint_likelihood}), the multi-sensor measurement pseudolikelihood assumes conditional independence between sensors resulting in a product likelihood form.
This means that if any subset of measurements in the measurement tuple result in a low joint pseudolikelihood when paired together, then the multi-sensor measurement pseudolikelihood for the measurement tuple will also be low.
This is intuitive as it means that a subset of measurements in the measurement tuple are unlikely to have been generated from the same target.

We propose a gating procedure by pre-computing (or approximating) the value for $\bar{\psi}^{(j^{(a)}, j^{(b)})}_Z$ for every $2$-pair measurement combination of sensors $a$ and $b$.
The value of the $2$-pair average pseudolikelihood can be checked against a gating threshold $\tau_G$ and stored in a $4-D$ matrix that is indexed by the pair of sensors and measurement indices, $M[a, b, j^{(a)}, j^{(b)}] = (\bar{\psi}^{(j^{(a)}, j^{(b)})}_Z < \tau_G)$.

When computing the transition density for $j^{(a)}$, an efficient check can be done to see if $M[a, b, j^{(a)}, j^{(b)}] = 0$ for any $j^{(b)} \in J^{-a}$.
If this occurs, then the transition to $j^{(a)}$ can be skipped by setting $p(j^{(a)}|J^{-a}) \approx 0$ without computing $\bar{\psi}^J_Z$.

The gating check is an $O(V)$ procedure, but maintains the overall complexity as quadratic in the number of sensors for both the Monte Carlo and Gaussian solutions in~\cite{Trezza2022}.
In the Monte Carlo approach, the value $\bar{\psi}^{(j^{(a)}, j^{(b)})}_Z$ can be pre-computed in $O(N_p)$ time resulting in a complexity of $O(N_pV^2m^2)$ to make the gate matrix $M$.


Alternatively, a Mahalanobis or Euclidean distance gating procedure between the maximum likelihood estimates $\hat{\textbf{x}}^{(a)}$ and $\hat{\textbf{x}}^{(b)}$ of $g^{(a)}(z^{(a)}_{j^{(a)}} | \textbf{x})$ and $g^{(b)}(z^{(b)}_{j^{(b)}} | \textbf{x})$ could be used as an approximation.
The Mahalanobis gate is constructed as $M[a, b, j^{(a)}, j^{(b)}] = (d^{2,(a)}_M(h^{(a)}(\hat{x}^{(b)})) < \tau_{G,a}) \lor (d^{2, (b)}_M(h^{(b)}(\hat{x}^{(a)})) < \tau_{G,b})$ where $d^{2,(a)}_M$ is the squared Mahalanobis distance for the distribution $g^{(a)}(h^{(a)}(\hat{x}^{(b)}))$ and similar for $d^{2, (b)}$.
If $g^{(s)}(\cdot)$ is Gaussian, then $d_M^{2, (s)}$ follows a chi-squared distribution with $n^{(s)}_Z$ degrees of freedom and the gate threshold $\tau_{G,s}$ can be set to achieve a particular gating probability using the cumulative chi-squared distribution.
The Euclidean distance gate is constructed as $M[a, b, j^{(a)}, j^{(b)}] = ||\hat{\textbf{x}}^{(a)} - \hat{\textbf{x}}^{(b)}||_2 < \tau_G$.
These simple checks often prove to be useful in many practical applications, especially when the measurement uncertainty is low.

\subsection{Memoization}\label{sec::proposed::memoize}


In memoization, we exploit the fact that many computations are repeated during the Gibbs sampling procedure and the results can be efficiently stored and accessed when needed.

A Gibbs sampler is a Markov chain Monte Carlo method that indirectly generates samples from a target distribution by simulating state observations of a Markov chain according to a stationary transition density.
This approach can be seen as a type of random walk through sample space, generating samples as it goes along.
At every step in the Markov chain simulation, the transition probabilities must be computed so the next chain observation can be sampled.
As the Markov chain simulation progresses, it is possible (and often likely) that the same state may be reached multiple times.
When this happens, the transition probabilities are unnecessarily recomputed each time, since the transition density is stationary.

In the multi-sensor adaptive birth sampler, recomputing the transition density can be computationally expensive.
Specifically, by Equation~(\ref{eq::orig_sampling_distr}), the average pseudolikelihood $\bar{\psi}^{J}_{Z}$ must be recalculated.
Note the same average pseudolikelihood will need to be computed every time a state wants to consider transitioning \textit{to} the state $J$, even if $J$ is never sampled.
This means that not only is the entire transition density for a given state unnecessarily recomputed, but the average pseudolikelihood will be unnecessarily recomputed by any state that considers transitioning into a state for which we've already computed $\bar{\psi}^J_Z$.
This can be costly, especially if this happens many times throughout the sampling procedure.

Further, after the Gibbs sampling truncation procedure is completed and a truncated label space is sampled, a component in the \ac{LMB} is then constructed for every newborn label.
By Equation~(\ref{eq::spatial_distr}) and Equation~(\ref{eq::birth_prob_hat}), to construct the labeled Bernoulli component the average pseudolikelihood (which was already calculated in the sampling procedure) must again be recalculated two more times.

As is commonly done in dynamic programming algorithms, we can avoid these redundant calculations by caching the value of $\bar{\psi}^J_Z$ once we compute it, and access it if it is needed again.
This can easily be implemented via a hash map keyed on the measurement tuple $J$, resulting in an $O(1)$ access operation.

\subsection{Prune and Cap Birth LMB}\label{sec::proposed::prunecap}
Although the Gibbs sampling procedure is designed to truncate many of the components with low birth probability, it is not guaranteed that all labels in the truncated \ac{LMB} will have a high birth probability.
Low birth probability labels may be sampled at any point, but will commonly occur in Markov Chain Monte Carlo methods while the sampler is converging to the target distribution.
In the $\delta$-\ac{GLMB} and \ac{LMB} filtering recursions, it is standard practice to prune and cap the belief state to remove low existence probability labels and cap the number of tracks kept alive to a manageable number \cite{Reuter2014}.
We can apply the same prune and cap procedure to the birth \ac{LMB} before providing them to the filter to reduce unnecessary computations.


\subsection{Missed Detection Sample Skipping}\label{sec::proposed::sampleskipping}
In missed detection sample skipping, we show that in many practical applications, such as those with ambiguous sensing modalities, if a newborn label is constructed by a multi-sensor measurement tuple with many missed detections, then it can lead to ambiguities in the downstream filtering recursion.
To prevent that, these sampled labels can often be skipped to reduce several unnecessary computations in the downstream filtering recursion.

Grouping the missed detected and detected elements of $J$ in $\psi^J_Z$ results in,
\begin{multline}\label{eq::psiJ_expand}
    \psi^J_Z(\textbf{x}) =
    \left[\prod\limits^V_{\substack{s'=1\\j^{(s') = 0}}} (1-p^{(s')}_D(\textbf{x}))\right]\\
    \times \left[\prod\limits^V_{\substack{s'=1\\j^{(s') > 0}}}\frac{p^{(s')}_D(\textbf{x}) g^{(s')}(z^{(s')}_{j^{(s')}}|x)}{\kappa^{(s)}(z^{(s')}_{j^{(s')}})}\right].
\end{multline}
When $p_B(x, l_+)$ and $p^{(s)}_D(\textbf{x})$ are not a function of $x$, then by Equation~(\ref{eq::spatial_distr}) the spatial distribution $p_B(x, l_+ | Z_J)$ constructed from a tuple $J$ with many missed detections is often uninformative.
This is specifically important when fusing ambiguous sensor measurements without an invertible observation function such as bistatic range, angle-only, \ac{TDOA} and \ac{FDOA} \cite{Murray2023}.
In these applications, a minimum number of observations from disperse sensors are often required to unambiguously compute an intersection point in cartesian space (e.g., angle triangulation).
For example in a multi-sensor 2D bearing-only sensing application, at least 2 sensors need to observe a target for it to be triangulated since range is unobserved.

In \cite{Trezza2022,Murray2023}, procedures such as triangulation do not explicitly occur as the ambiguity naturally resolves through the multi-sensor pseudolikelihood function in Equation~(\ref{eq::spatial_distr}).
However, if the number of non-missed detection observations in $J$ is below the minimum number of observations required to resolve this ambiguity, the spatial distribution will likely be uninformative and may falsely associate with measurements in the $\delta$-\ac{GLMB} update procedure.
For these types of applications, it is often useful to make an explicit rule on the minimum number of non-missed detection measurements in $J$ for a newborn component to be constructed.

Similarly, in applications containing one or more sensors with a high clutter rate, it is likely that a state containing a clutter measurement may be transitioned to during the Gibbs sampling procedure.
Since a target did not generate the clutter measurement it is often most probable that the remaining sensors missed the detection, resulting in sampling the all-but-one missed detection measurement tuple $J = (-1,\dots, j^{(s)}, \dots, -1)$.
In this case, a labeled Bernoulli component in the \ac{LMB} would be constructed from this clutter observation.

In both cases, the resulting birth probability of the constructed labeled Bernoulli component will likely be low if one or more sensors that missed the detection have a high detection probability.
However, as discussed in Section~\ref{sec::proposed::prunecap}, by adding this component in the birth \ac{LMB} the onus is now shifted to the $\delta$-\ac{GLMB} to truncate any hypotheses containing this label, resulting in many unnecessary computations that could be avoided by simply pruning components from the birth \ac{LMB} that were constructed with a tuple containing more than a maximum number of missed detections.
This procedure will often coincide with the pruning procedure discussed in Section~\ref{sec::proposed::prunecap} but can be valuable to check explicitly.

\section{Simulations}\label{sec::sim}

\begin{figure}[t]
    \includegraphics[width=0.48\textwidth, height=6cm]{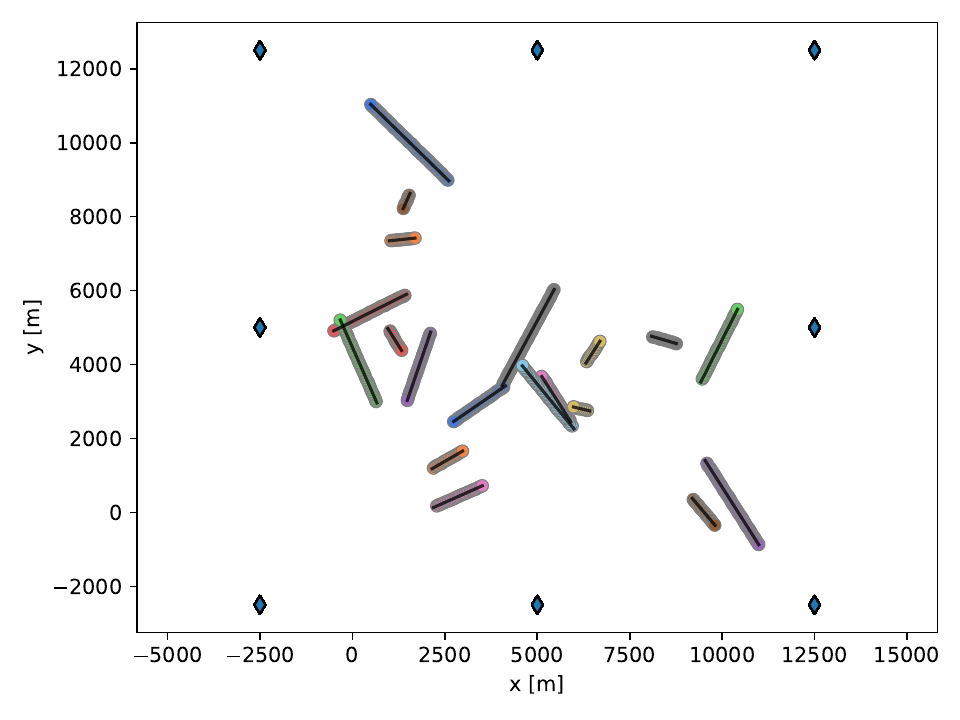}
    \caption{Single observation of target trajectories (black lines), and an example labeled state estimate result (colored circles). Diamonds represent the locations of the bearing-range sensors.}\label{fig::qual}
    \label{fig::laydown}
\end{figure}

In this section, we used a similar simulation as Scenario 1 from~\cite[Section VIII.A]{Trezza2022} with the particle \ac{LMB} filter to highlight the benefits of the proposed efficiency improvements.
The target and sensor positions of the simulation are shown in Figure \ref{fig::laydown}.
The simulations were implemented in Python and run on an AMD EPYC 7742 64-Core Processor with 16GB of RAM.
The runtime results include other inefficiencies such as logging, but are held constant throughout the comparison.
Additional runtime improvements would be expected if the approaches were implemented in a compiled language such as C/C++, but the relative differences between the techniques would be expected to be similar.

The scenario tracks a time varying, randomly generated number of targets, using 8 bearing-range sensors.
The targets are 2D planar position and velocity, $x = [p_x, \dot{p}_x, p_y, \dot{p}_y]^T$.
Bearing-range measurements, $z^{(s)} = [\alpha^{(s)}, r^{(s)}]^T$, were observed according to the single-target measurement likelihood $g(z^{(s)}|x) = \mathcal{N}(z^{(s)}; h^{(s)}(z^{(s)}, x^{(s)}), R^{(s)})$,
where $x^{(s)} = [p^{(s)}_x, p^{(s)}_y]^T$, $R^{(s)} = \text{diag}(0.25^2, 10^2)$, and
\begin{align}
    h_\alpha^{(s)}(x, x^{(s)}) &= \arctan\left(\frac{p^{(s)}_x - p_x}{p^{(s)}_y - p_y}\right)\\
    h_r^{(s)}(x, x^{(s)}) &= \sqrt{(p^{(s)}_x - p_x)^2 + (p^{(s)}_y - p_y)^2}.
\end{align}

The number of targets birthed was sampled uniformly every 5 seconds, $0-3$ targets were selected to be born.
Target positions were uniformly sampled over the domain $[0, 10000]\;m$ and target speed was fixed at $50\;m/s$.
Target velocity heading was uniformly sampled in the domain $[-\pi, \pi]$ radians.

The target dynamics followed a nearly constant velocity transition model $f_+(x_+|x) = \mathcal{N}(x_+; (I_2 \otimes F)x, (I_2 \otimes G)w)$ with $I_2$ as the $2\times2$ identity matrix, $\otimes$ being the Kronecker product, and
\begin{equation*}
    F =
    \begin{bmatrix}
        1 & \Delta_t \\
        0 & 1
    \end{bmatrix},\qquad
    G =
    \begin{bmatrix}
        \frac{\Delta_t^2}{2} \\
        \Delta_t
    \end{bmatrix},
\end{equation*}
where $\Delta_t$ was the discrete-time sampling interval \cite{Li2003}.
The acceleration white noise was $w = [5, 5]^T\;m/s^2$.
The simulations ran for 100 seconds with $\Delta_t = 1$ second.
Each target had a survival probability of $p_s(x,l) = 0.99$.
The sensors used a constant detection probability of $p_D^{(s)}(x,l) = 0.95$.
Clutter was modeled as Poisson distributed with intensity $\kappa^{(s)}(\mathbb{Z}^{(s)}) = \lambda_c^{(s)} \mathcal{U}(\mathbb{Z}^{(s)})$ where $\mathcal{U}(\mathbb{Z}^{(s)})$ is the uniform distribution over $\mathbb{Z}^{(s)}$, and $\lambda^{(s)}_c = 10$.

The adaptive birth Gibbs sampler used the multi-short run sample generation architecture as described in~\cite{Trezza2023}, with 20 chains and a chain length of 5.
The initial condition for all chain simulations was the all-miss-detected measurement tuple.
The maximum birth probability was $r_{B, max} = 1.0$, and the expected birth rate was $\lambda_{B,+} = 0.5$.
The maximum association probability was $\tau = 0.01$.

\subsection{Results}\label{sec::sim::results}

The overall runtime reductions are shown in Table~\ref{table::runtimes}.
Each method was run individually and compared to the baseline run time, and one simulation was included where all efficiency improvements were combined and compared to the baseline.
Run times were recorded using wall clock time.
Each percent reduction was calculated as $ \frac{r_{1} - r_{2}}{r_{1}} * 100$, where $r_{1}$ is the baseline simulation's runtime and $r_{2}$ is runtime of interest.

\begin{figure}[t]
    \includegraphics[width=0.48\textwidth, height=6cm]{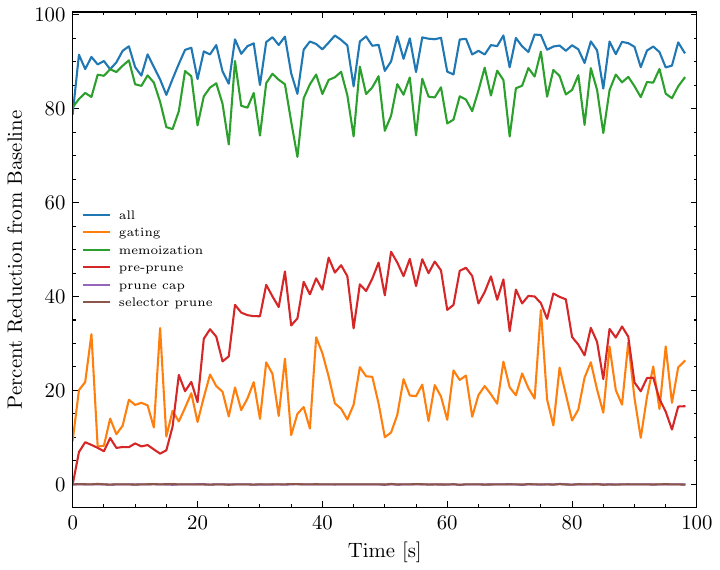}
    \newline
    \caption{Percent change (decrease) in number of evaluations per method compared to baseline.}
    \label{fig::evals}
\end{figure}

\begin{table}[]
    \centering
    \caption{Percent reduction in runtime compared to baseline.}
    \begin{tabular}{|l|l|l|l|}
        \hline
        \textbf{Feature}         & \textbf{ Reduction (\%)} \\ \hline
        Pre-prune       & 33.45          \\ \hline
        Gating          & 11.17          \\ \hline
        Memoization     & 83.82          \\ \hline
        Prune and Cap   & 2.09           \\ \hline
        Sample skipping & 0.90           \\ \hline
        All on          & 91.57          \\ \hline
    \end{tabular}
    \label{table::runtimes}
\end{table}

\subsubsection{Associated Measurement Pre-pruning}\label{sec::sim::preprune}

In this simulation, we used a small association threshold of $\tau_A = 0.0001$ to demonstrate the effectiveness of the approach when only using measurements that are highly likely to be unassociated with existing targets.
In practice, this value can be increased to allow for birthing from more ambiguous measurements if necessary.

The measurement pre-pruning method prunes already associated measurements, once track labels are established.
This resulted in an average of 4,530 values of $\bar{\psi}^J_Z$ per timestep that did not need to be computed, which is a 31.4\% average reduction in number of evaluations, as shown in Figure~\ref{fig::evals}. 
Overall, this translated to a runtime reduction of 33.45\%, as shown in Table~\ref{table::runtimes}.

\subsubsection{Measurement Gating}\label{sec::sim::gate}

For this simulation, we used a very simple Euclidean distance coarse gate with $\tau_G = 500 m$ to highlight the benefit of this procedure.
On average, this resulted in 2,588 values of $\bar{\psi}^J_Z$ that did not need to be computed at each timestep, an average reduction in the number of evaluations by 19.06\%, as shown in Figure~\ref{fig::evals}. 
Including the time it took to construct the gate matrix, this translated to a total runtime reduction of 11.17\%, as shown in Table~\ref{table::runtimes}.

\subsubsection{Memoization}\label{sec::sim::memoize}

Memoization resulted in the most significant reduction in runtime from any single proposed method.
The birth Gibbs sampler explores many repeated solutions, so saving these results instead of recalculating them produces an 83.82\% reduction in runtime, as shown in Table~\ref{table::runtimes}. 
For this simulation, an average of 11,275 $\bar{\psi}^J_Z$ evaluations were saved per timestep, an 83.53\% reduction in total $\bar{\psi}^J_Z$ evaluations, as shown in Figure~\ref{fig::evals}.  

\subsubsection{Prune and Cap Birth LMB}\label{sec::sim::prunecap}

To prune, we drop any track labels with an $r_{l+} < 0.001$.
We cap our track label cardinality to 100.

Pruning and capping resulted in a 2.09\% reduction in runtime, as shown in Table~\ref{table::runtimes}.
This shows there are some savings in this scenario, but we would see a more significant runtime reduction in a scenario that require more Gibbs samples. 

\subsubsection{Missed Detection Sample Skipping}\label{sec::sim::sampleskipping}

In this simulation, a maximum number of missed detections was set to 4.

\begin{figure}[t]
    \includegraphics[width=0.48\textwidth, height=6cm]{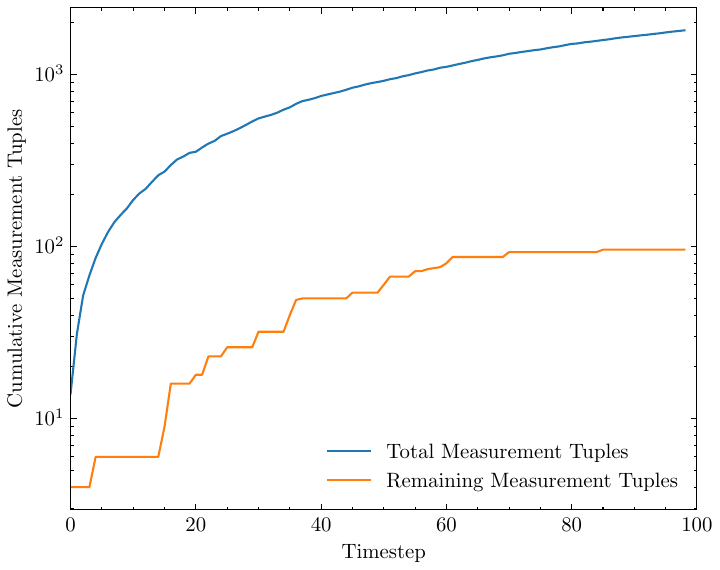}
    \newline
    \caption{Measurement tuples skipped when in labeled Bernoulli components construction.}
    \label{fig::selectorprune}
\end{figure}

As seen in Figure~\ref{fig::selectorprune}, a significant number of measurement tuples were skipped even for this sensing application with unambiguous measurement inverse functions and a relatively low clutter rate.
Despite the large number of skipped components, this only translated to a total runtime reduction of 0.90\%, as shown in Table~\ref{table::runtimes}.
This is because the measurement tuples themselves were still considered in the sampling procedure, meaning $\bar{\psi}^J_Z$ was computed every time it was evaluated by the Gibbs sampler.

\subsubsection{Overall Reduction}\label{sec::sim::allon}

This paper focuses on adaptive birth efficiency improvements, in our baseline simulation the birth procedure was 98\% of the runtime.
After combining the above techniques to improve the runtime of the birth procedure, we see an overall simulation runtime reduction of 91.57\%, with an average of 91.54\% reduction in evaluations, as shown in Figure~\ref{fig::evals}. 

\begin{figure}[t]
    \includegraphics[width=0.48\textwidth, height=6cm]{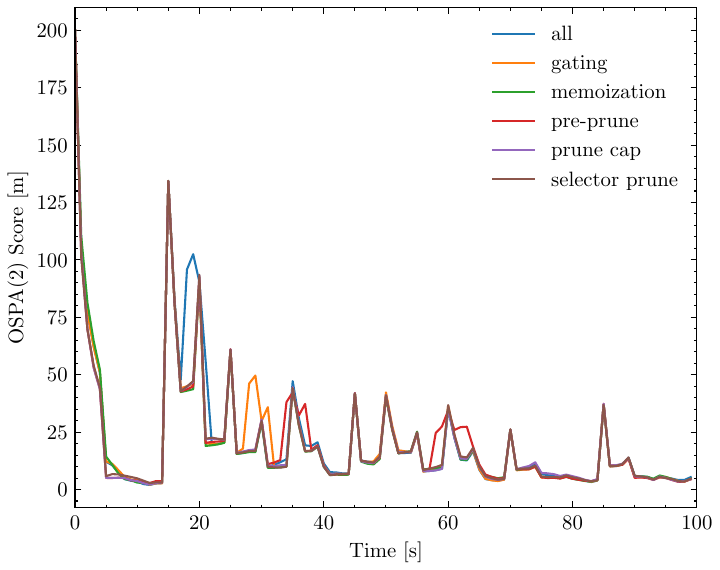}
    \newline
    \caption{Negligible variation in OSPA(2) performance with efficiency suggestions implemented.}
    \label{fig::ospa}
\end{figure}

To verify these efficiency improvements don't affect tracking performance, we use the \ac{OSPA}(2) \cite{Beard2020} to compare results, as shown in Figure \ref{fig::ospa}.
The \ac{OSPA}(2) metric was computed using a distance cutoff of $200$, a distance order of $1.0$, a sliding window length of $5$ and an expanding window weight power of $0$.
There is some small variation in tracking performance with each proposed method, but this is expected because each method modifies or truncates different components of the birth procedure and this is not a Monte Carlo analysis.
The overall tracking performance stayed consistent, while the runtime is significantly reduced.

\section{Conclusion}\label{sec::conclusions}

This paper provided five methods for improving the runtime of the multi-sensor measurement adaptive birth procedure.
There are two main components of the adaptive birth procedure where we focused on reducing runtime.
The first was the Gibbs sampler, which is a large driver of the high runtime.
The second is during construction of the truncated birth LMB, which is a smaller proportion of computations, but can still result in reducing overall runtime.
Some methods are more application specific and would see more improvement in scenarios with less accurate sensors, or scenarios with more clutter.
Overall these efficiency improvements significantly reduce runtime in a variety of scenarios.
This paper provided benchmarking for what method to apply, depending on where the birth procedure bottleneck is in a given scenario.

\bibliographystyle{IEEEtran}
\bibliography{IEEEabrv, ms.bib}

\end{document}